\documentclass[journal=nalefd,manuscript=letter]{achemso}

\usepackage{graphicx}
\usepackage{color}
\usepackage[latin1]{inputenc}
\usepackage{amsbsy,amsmath,amssymb,amsthm,amsfonts}
\usepackage{times,mathptmx}
\usepackage{bm}
\usepackage{units}
\newcommand{\comment}[1]{}
\newcommand{\eref}{Eq.\eqref}

\definecolor{Blue2}{rgb}{0,0.0,0.7}
\definecolor{Red2}{rgb}{0.7,0.0,0.}
\definecolor{Green2}{rgb}{0.0,0.5,0.}
\definecolor{darkblue}{rgb}{0,0,.5}

\author{Torben Winzer}
\author{Andreas Knorr}
\author{Ermin Malic}
\email{ermin.malic@tu-berlin.de}
\affiliation{Institut f{\"u}r Theoretische
Physik,  Technische
Universit{\"a}t Berlin, 10623 Berlin, Germany}

\title
{Carrier Multiplication in Graphene}

\begin{document}
\begin{abstract}
Graphene as a zero-bandgap semiconductor is an ideal model structure
to study the carrier relaxation channels, which are inefficient in
conventional semiconductors. In particular, it is of fundamental interest to
address the question whether
 Auger-type processes significantly influence the carrier dynamics in graphene.
These scattering channels bridge the valence and conduction band
allowing  carrier multiplication - a process that generates multiple charge carriers from the absorption of a single photon.
This has been suggested
in literature for improving the efficiency of solar cells. Here we
show, based on microscopic calculations within the density matrix
formalism, that Auger processes do play an unusually strong role for
 the relaxation dynamics
of photo-excited charge carriers in graphene. We predict that a considerable carrier multiplication takes place, suggesting graphene as a 
new material for
high-efficiency solar cells and for high-sensitivity photodetectors.
\end{abstract}


\begin{figure}[b!]
\center{\includegraphics[width=0.5\columnwidth]{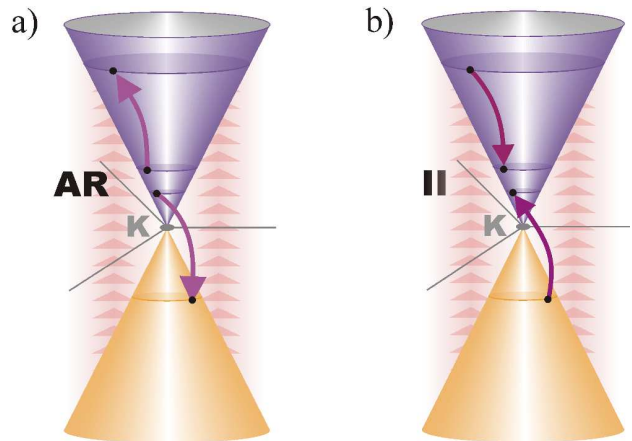}
 }
 \caption{Linear energy
dispersion of graphene around the $K$ point. After an optical
excitation (depicted by red arrows), the hot carriers relax toward equilibrium via
Coulomb-induced scattering processes. The figure illustrates the two
Auger-type relaxation channels: a) Auger recombination (AR) and b)
inverse Auger recombination or impact ionization (II).
}\label{auger}
\end{figure}

\vspace{12pt}
Graphene is a strictly two-dimensional zero-bandgap semiconductor with a linear
energy dispersion. It has exceptional optical and electronic properties, which
have sparked interest in both fundamental research and industry \cite{novo07, neto09, geim09}. The key for designing
 and engineering novel graphene-based
optoelectronic devices is a microscopic understanding of the
ultrafast relaxation dynamics of non-equilibrium carriers. To give
an example, short relaxation times are crucial for realization
high-performance graphene-based saturable absorbers \cite{zhang09}.
A further challenging feature of graphene relaxation dynamics is the expected high efficiency of 
Auger relaxation, which  can be exploited to
obtain carrier multiplication - a process that has been
intensively discussed for improving the efficiency of solar energy conversion \cite{nozik09, nozik02, klimov08, klimov09}. Here, multiple charge carriers are generated from the absorption of a single photon. We distinguish two types of Auger
processes:  Auger recombination (AR) and  impact ionization (II),
cp.   \ref{auger}. AR is a process, where an electron is
scattered from the conduction into the valence band, while at the
same time, the energy is transferred to another electron, which is
excited to an energetically higher state within the conduction band (\ref{auger}a).
II is the opposite process (inverse Auger recombination): An
electron relaxes to an energetically lower state inducing the
excitation of a valence band electron into the conduction band (\ref{auger}b). The
result of II is an increase of the carrier density (carrier multiplication). Both processes  also
occur for holes in an analogous way. In conventional semiconductor
structures, these relaxation channels are suppressed by restrictions
 imposed by energy and momentum conservation,
which are difficult to fulfill at the same time due to the bandgap and the energy dispersion.
In contrast, graphene  is expected to show a very efficient scattering via Auger processes. Here, it is 
of crucial importance to address the question whether
impact ionization is efficient enough
to give rise to a significant carrier multiplication despite the competing processes of Auger recombination,
 phonon-induced scattering, and intra-band relaxation.

Only few experimental studies on ultrafast relaxation dynamics in
graphene have been reported \cite{dawlaty08, sun08, elsaesser09}.
Common to all investigations is the observation of two distinct time
scales in differential transmission spectra: A fast initial decay of
the pump-induced transmission on a timescale of some tens of
femtoseconds and a slower relaxation process in  the range of some
hundred femtoseconds. The fast decay is ascribed to Coulomb-induced
carrier scattering, while the slower process is associated with
carrier cooling due to electron-phonon scattering. The experimental
data has not yet been complemented by theoretical studies treating all relaxation channels on a consistent microscopic footing. In particular, to best
of our knowledge there have been no investigations on the Coulomb-induced fast decay component so far.
Previously published reports on the time-resolved relaxation
dynamics in graphene focus on the relaxation channels via optical or
acoustic phonons accounting for the energy dissipation of
photo-excited electrons \cite{butscher07, bistritzer09, tse09}.
Others investigated carrier generation and recombination rates based
on a static theory \cite{rana07}. Our microscopic approach, presented in this work,  allows a time-resolved study of Auger processes leading to new insights
on their efficiency as a function of time starting from the initial non-equilibrium situation up to the achievement of an equilibrium.

In this article, we present a microscopic study of the Coulomb- and phonon-induced
relaxation of photo-excited carriers in graphene.  Our approach is
based on the density matrix formalism describing the coupled
population and coherence dynamics including Pauli blocking terms. 
The starting point for our investigations is the Hamilton operator
$H$ including the non-interacting contribution of carriers and
phonons, the electron-light coupling, the Coulomb interaction, and
the electron-phonon interaction: $
  H={\color{Green2}{H_0}}+{\color{Red2}{H_{\text{carrier-light}}}}+{\color{Blue2}{H_{\text{Coulomb}}}}+H_{\text{el-ph}}.
$
With the above Hamilton operator, we can obtain the temporal evolution of the density matrix elements via Heisenberg equation 
\cite{lindberg88, haug, rossi02, kira06} yielding
microscopic many-particle
graphene Bloch equations:
\begin{eqnarray}
\label{bloch1}
 \dot{f}^{\lambda}_{\bm{k}}(t) &=&
{\color{Red2}{{2Im\left(\Omega^* p_{\bm{k}} \right)}}}+
{\color{Blue2}{\Gamma_{\bm{k},\lambda}^{in}\left(1-f^{\lambda
}_{\bm{k}} \right)-\Gamma_{\bm{k}, \lambda}^{out} f^{\lambda
}_{ \bm{k}}}}\,, \\[5pt]
\label{bloch2}
\dot{p}_{\bm{k}}(t)&=&{\color{Green2}{-i\Delta\omega_{\bm{k}}p_{\bm{k}}}}
{\color{Red2}{-i \Omega \left(f^{c}_{\bm{k}} -f^{v}_{\bm{k}}
\right)}}{\color{Blue2}{ -\gamma_{\bm k} p_{\bm k}+\sum_{\bm k'}\gamma_{\bm
k'}p_{\bm k'}}}\,,
\end{eqnarray}
 with the carrier population $f^{\lambda
}_{\bm k}$ in the conduction ($\lambda=c$) and the valence
($\lambda=v$) band \cite{haug}. The microscopic polarization $p_{\bm k}$ is a measure for the transition probability. Different terms in
the the graphene Bloch equations (color-coded) stem from different Hamilton operator
contributions:
\\(i) The free carrier part ${\color{Green2}{H_0}}$ contains the energy dispersion, which is linear around the
$K$ and $K'$ points,
 $\hbar\omega_{\lambda \bm{k}}=\pm\hbar v_F|\bm{k}|$, with
the electron velocity $v_F=\unit[10^6]{ms^{-1}}$ and $\bm k$ as the wave vector, see
  \ref{auger}. It leads to the first term in \eref{bloch2} with
the energy difference between the conduction $(\lambda=c)$ and the valence band 
$(\lambda=v)$ $\hbar\Delta\omega_{\bm{k}}=\hbar(\omega_{c\bm{k}}-\omega_{v\bm{k}})$.\\
(ii) The carrier-light interaction
${\color{Red2}{H_{\text{carrier-light}}}}$ leads to the Rabi
frequency $\Omega(t)$, which is determined by the the
vector potential and the optical matrix element \cite{malic06b}. The excitation field, which optically transfers
electrons from the valence band to the conduction band, is described
by a Gaussian pulse with the width
$\sigma=\unit[10]{fs}$ and  the excitation energy $\hbar
\omega_L=\unit[1.5]{eV}$ (chosen in agreement with a recent
experimental realization)\cite{elsaesser09}. \\
(iii) The Coulomb interaction ${\color{Blue2}{H_{\text{Coulomb}}}}$
induces scattering between carriers leading to the relaxation of
photo-excited electrons and holes.
The investigation of  non-equilibrium processes requires a description
beyond the Hartree-Fock level. We treat the Coulomb interaction up to
the second
order Born-Markov approximation \cite{haug} yielding the
Boltzmann-like scattering contributions in equation  (\ref{bloch1}).
The  time-dependent Coulomb in- and out-scattering  rates $\Gamma_{\bm{k}
\lambda}^{in/out}(t)$ are
determined microscopically. They explicitly contain intra- 
and intervalley 
as well as intra-
and interband scattering processes, which  fulfill the momentum and the energy conservation.
The  strength of our approach is the possibility to access the time-, momentum-, and
angle-resolved relaxation dynamics of non-equilibrium carriers. 
For our investigations, we
take all relaxation paths into account focusing in particular on
Auger-type processes. Electron-phonon scattering - a competing relaxation channel, is treated on the same microscopic footing.\\

With all ingredients at hand, we can resolve the relaxation
dynamics of photo-excited carriers in graphene: First, electrons are
optically excited from the valence into the conduction band by
applying a \unit[10]{fs} laser pulse with an energy of
$\unit[1.5]{eV}$ resulting in a non-equilibrium distribution. The hot charge carriers relax 
toward equilibrium
via Coulomb and phonon-induced scattering processes. For a strong
optical excitation, we observe  thermalization of  carriers within the first hundred femtoseconds followed by 
carrier cooling induced by electron-phonon scattering. Our results are in
agreement with the two decay components observed in
experimental differential transmission spectra \cite{dawlaty08,
elsaesser09}.
Furthermore, our investigations show that the intervalley processes play a minor role for the 
Coulomb-induced relaxation dynamics in graphene.
This can be explained by the large momentum transfer, which is
necessary to scatter electrons between the valleys. Since the Coulomb matrix element decreases
with the momentum transfer, these processes have a negligible contribution. We have
also studied the influence of the momentum angle between two
scattering electrons. Similar to electron-phonon coupling, where
the scattering is maximal for carriers with parallel momenta
\cite{butscher07}, electron-electron scattering also shows a pronounced
angle-dependence.
\\
\begin{figure}[t!]
\center{\includegraphics[width=0.75\columnwidth]{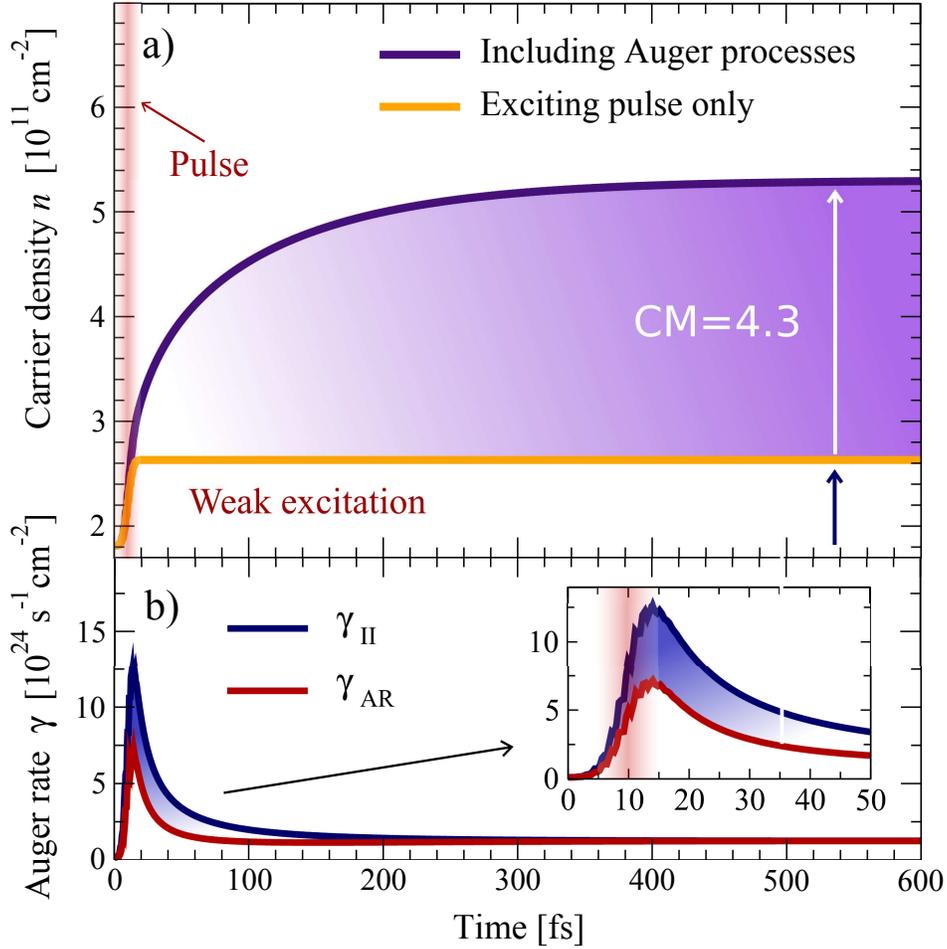} } 
\caption{a) Temporal
evolution of the charge carrier density $n$ (electrons in conduction
and holes in valence band) for a weak exciting pulse (its width is indicated by the red-shaded area) inducing
initial carrier densities in the range of $\unit[10^{11}]{cm^{-2}}$.  
The
figure illustrates the significance of impact ionization leading to
 carrier multiplication (CM) by a factor of two, i.e. the carrier density induced by the optical excitation 
(orange line) is doubled during the relaxation process. b) Rates for
impact ionization (II) and Auger recombination (AR) as a function of
time. The figure illustrates the temporally broad asymmetry between
these two Auger processes in favor of II.}\label{auger_weak}
\end{figure}

Now we can also answer the question whether
Auger-type processes play a significant role for the relaxation
dynamics in graphene:
\ref{auger_weak}a shows
the temporal evolution of the carrier density after an optical
excitation. Neglecting all interactions, we can first model the influence of the exciting pulse (red-shaded area in   \ref{auger_weak} illustrates
 its width of \unit[10]{fs}). It generates electrons in conduction and holes in valence band leading to an increase of the charge carrier density $n$, as long as the 
pulse is present (orange line).
 The inclusion of
Coulomb-induced carrier scattering leads to a significant increase
of the carrier density, even after the pulse is switched off (purple
line). This process - generation of multiple charge carriers by absorption of a single photon - is called  carrier multiplication (CM). It is 
illustrated by the purple-shaded
area in    \ref{auger_weak}a. This can be traced back to the process of
impact ionization, where valence band electrons are excited into the conduction band, cp.   \ref{auger}b. 
To prove that our interpretation is correct, we
switched off all Auger contributions to the relaxation dynamics. We
obtain constant carrier densities after the influence of the pulse.
As a conclusion, Auger processes must be responsible for the strong increase of the charge carrier density.

The observed carrier multiplication by a factor of approximately four can be explained by an
asymmetry between impact ionization (II) and  Auger recombination (AR) resulting in a much higher probability
for II. Otherwise, the created carriers would scatter back into the
valence band via AR with the same probability and the carrier multiplication would not
occur. The asymmetry can be explained as follows:
The squares of the corresponding matrix elements entering
the scattering rates $\Gamma_{\bm{k}
\lambda}^{in/out}(t)$ in \eref{bloch1} are equal, but the densities of  final states for AR and II
are quite different, cp. also  Ref. \cite{zunger06}. To give an
example, the probability for an electron to be excited into the
conduction band is proportional to $\text{II} \propto
f_k^{v}(1-f_k^{c})$, while the opposite process is $\text{AR}
\propto f_k^{c}(1-f_k^{v})$. In the first femtoseconds after the
optical excitation, the probability to find an electron (a hole) in
the conduction (valence) band close to the $K$ point is small, i.e.
$f_k^{c} \approx 0$ ($f_k^{v} \approx 1$). As a result, $\text{II}
\approx 1$ and $\text{AR} \approx 0$. In other words, at the
beginning of the relaxation dynamics, the Auger recombination is
suppressed by Pauli blocking, since its final states in the valence
band are occupied. With increasing relaxation time, an equilibrium
between II and AR is reached resulting in a constant carrier
density. This interpretation is confirmed by   \ref{auger_weak}b,
where the Auger rates $\gamma_{AR}$ and $\gamma_{II}$ are shown as a
function of time.
They
 are obtained by summing all scattering rates, which contribute to the processes of II and AR, respectively. 
\ref{auger_weak}b
illustrates the high efficiency of both Auger processes  with rates
around $\gamma=\unit[10^{24}]{s^{-1}cm^{-2}}$, which is a few orders
of magnitudes larger than in conventional GaAs semiconductor quantum
wells or InAs/GaAs quantum dots \cite{harrison}. In the first
$\unit[300]{fs}$ after the excitation, we find a large asymmetry
between II and AR with $\gamma_{II}>\gamma_{AR}$ explaining the
observed carrier multiplication in   \ref{auger_weak}a. When the carriers reach
equilibrium, the two rates approach the same value
$\gamma_{II}=\gamma_{AR}=\unit[2.5\cdot10^{24}]{s^{-1}cm^{-2}}$.
\begin{figure}[t!]
\center{\includegraphics[width=0.75\columnwidth]{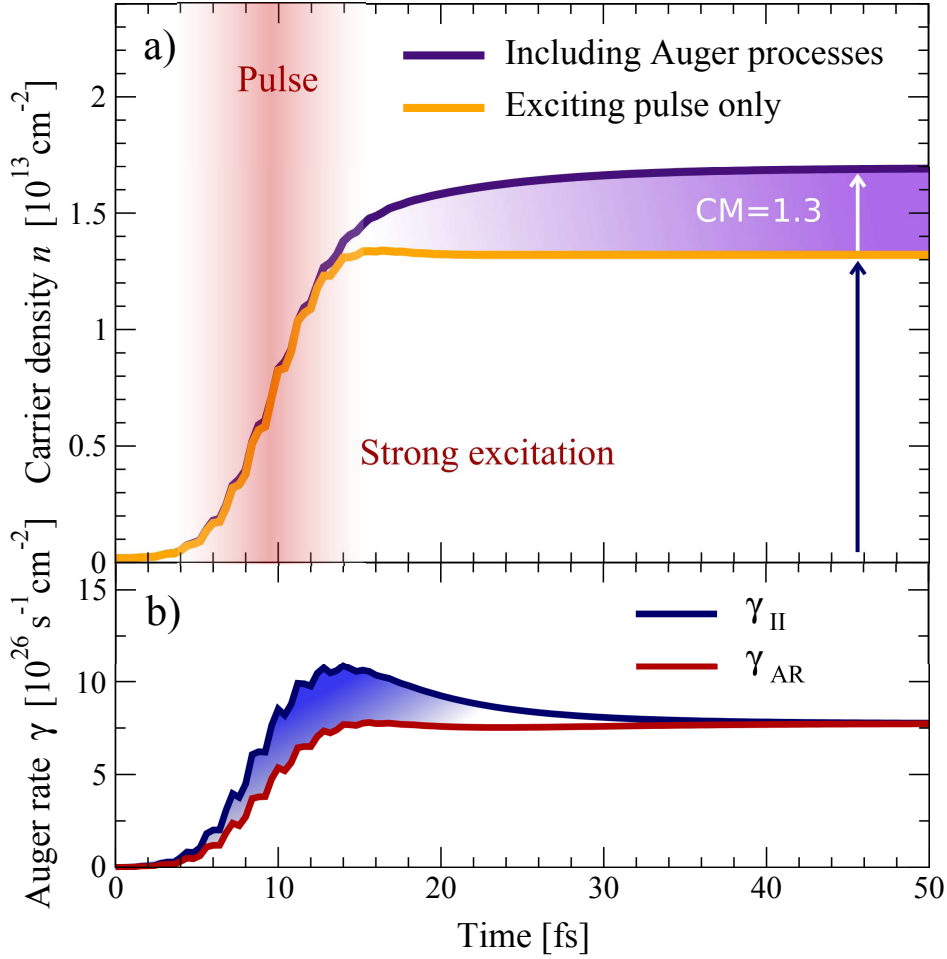} } \caption{The same as   \ref{auger_weak}, just for the case
of a strong excitation pulse, inducing initial carrier densities in the range of
$\unit[10^{13}]{cm^{-2}}$.  
Here, the carrier multiplication (CM) is
smaller than  in the case of weak excitation, cp.
  \ref{auger_weak}a, which can be explained by the temporally narrower asymmetry between the impact ionization and 
Auger recombination.}\label{auger_strong}
\end{figure}

The efficiency of carrier multiplication depends on the strength of the optical
excitation, i.e. on how many charge carriers are present in the
structure and how efficient Pauli blocking and the resulting
asymmetry between II and AR is. \ref{auger_strong} shows the
temporal evolution of the charge carrier density for a strong
excitation pulse inducing initial carrier densities in the range of
$\unit[10^{13}]{cm^{-2}}$. Here, the Coulomb-induced relaxation
dynamics is accelerated, since the number of scattering partners is
increased. Auger-type processes as well as
 intraband carrier-carrier processes
become very efficient leading to the thermalization of the
electronic system within few tens of femtoseconds, cp.
  \ref{auger_strong}b. Both II and AR rates are two orders of
magnitude larger than in the case of a weak optical pulse, cp.
  \ref{auger_weak}b. However, the asymmetry between II and AR is
only given for a narrow time slot, since at high carrier densities an equilibrium is reached
very fast. Already after approximately \unit[35]{fs}, the II and AR
rates are equal reaching the value
$\gamma_{II}=\gamma_{AR}=\unit[8\cdot10^{26}]{s^{-1}cm^{-2}}$. As a
result, carriers can be multiplied only during a very short time
period leading to a  carrier multiplication of 1.3, which is smaller than in
the case of weak excitation, cp.   \ref{auger_weak}a. The observed dependence on the exciting pulse intensity, i.e. the 
carrier density, is in agreement with a first experimental study \cite{george08}.\\

\begin{figure}[t!]
\center{
\includegraphics[width=0.75\columnwidth]{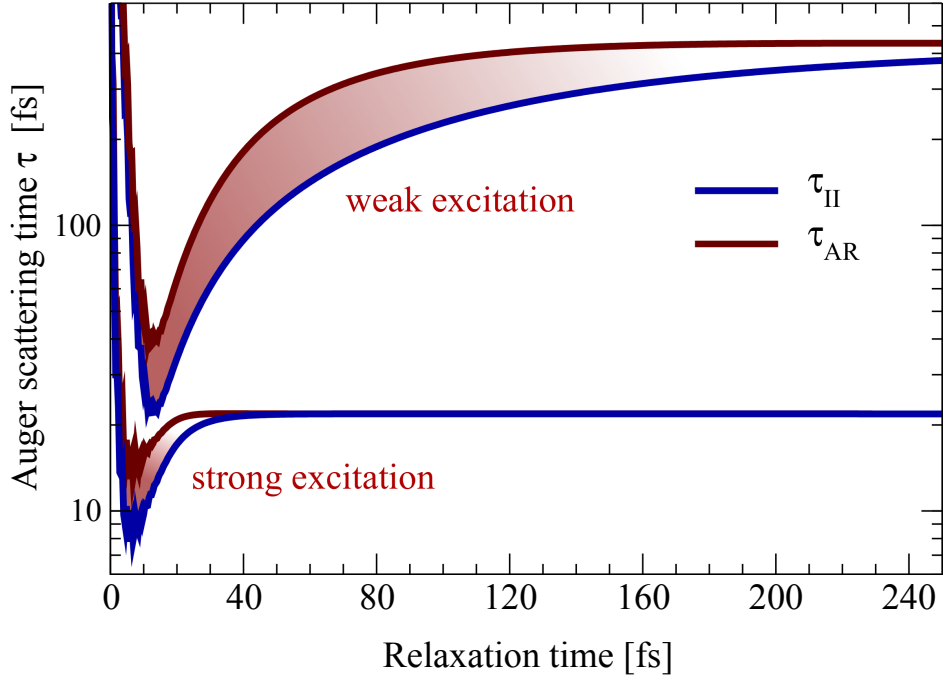} } \caption{Logarithmic illustration of
the asymmetry between the Auger recombination and impact ionization
by plotting the corresponding scattering times $\tau_{II}$ and
$\tau_{AR}$ during the relaxation process for both weak and strong
excitation. The region of asymmetry between II and AR (shaded area) is much larger and
temporally broader for weak excitation explaining the larger carrier multiplication
observed in this case, cp.   \ref{auger_weak}a. }\label{asym}
\end{figure}
Finally, to study the asymmetry between II and AR, we introduce the Auger
scattering time $\tau=n/\gamma$ with the time-dependent carrier
density $n$, cp.   \ref{asym}. During the first \unit[10]{fs}, the exciting pulse increases the number of charge carriers
resulting in efficient Auger scattering processes, which is reflected by a rapid decrease of the 
scattering time.
In the case of a weak (strong) excitation, the minimal time lies at $\tau_{II}\approx\unit[20]{fs}$ and
$\tau_{AR}\approx\unit[40]{fs}$ ($\tau_{II}\approx\unit[8]{fs}$, $\tau_{AR}\approx\unit[13]{fs}$). With increasing relaxation time,
the Auger processes become less efficient due to the enhanced Pauli blocking. In equilibrium, the
scattering times are $\tau_{II}\approx\unit[22]{fs}$ and
$\tau_{II}\approx\unit[425]{fs}$ for a strong and a weak exciting
pulse, respectively.
\ref{asym} illustrates that for weak excitation, i.e. small carrier densities,  the
asymmetry between II and AR is temporally broad offering enough time
for a considerable increase of charge carriers. In contrast, for
strong excitation II is more efficient only for a small time range
directly after the exciting pulse. This results in a smaller value
for carrier multiplication.  \\

To understand the efficiency of carrier multiplication, we need to take into account
electron-phonon scattering. This  is an important relaxation
channel, which is in direct competition with Auger-type processes.
Here, the excited electrons are cooled by emission of phonons
resulting in a loss of energy necessary for Auger processes. The
 inclusion
of electron-phonon
 coupling in graphene Bloch equations is straight-forward and yields contributions
 similar to the electron-electron scattering. For more details, see Ref. \cite{butscher07}. As expected, our calculations 
show that phonons reduce
the efficiency of carrier multiplication (not shown). However, it still
remains significant enough reflecting the strong Coulomb interaction in
graphene and in particular the large and temporally broad asymmetry
between II and AR. 
This finding implies graphene to be a promising candidate for high-sensitivity photodetectors and high-efficiency solar cells. 
It is beyond the scope of our work to discuss the difficulties one will be confronted with to realize such  a device, e.g. the fast 
extraction of 
generated carriers before they relax radiatively. 
We focus on microscopic investigations to give new insights in the temporal dynamics of Auger processes, in particular on the parameter range of
an efficient carrier multiplication.    

In summary, we have microscopically investigated the charge
carrier relaxation and multiplication  in graphene. In agreement with recent
experiments, we obtain a Coulomb-induced thermalization of the
strongly excited system within the first hundred femtoseconds followed by
a cooling of carriers via electron-phonon scattering. In particular,
we find Auger-type processes to have a significant influence on the
relaxation dynamics. We observe a strong asymmetry between impact
ionization and Auger recombination leading to a significant multiplication of
charge carriers. This process is found to be even more pronounced for very small optical excitations, such as solar radiation. 
Our calculations reveal that even for small bandgaps of up to $\unit[100]{meV}$ (modeling graphene nanoribbons), a 
considerable carrier multiplication still occurs. 
In conclusion, our fundamental investigations show that Auger-type processes might be 
of importance for the application of  graphene as a new material for high-efficiency solar cells and high-sensitivity photodetectors.   
\\

We acknowledge the support from SFB 658  and 
GRK 1558.
Furthermore, we thank U. Woggon (TU Berlin) for discussions on carrier multiplication as well as T. Elsaesser, M. Breusing (MBI Berlin), and
F. Milde (TU Berlin) for discussions on carrier dynamics in graphene.
\providecommand*{\mcitethebibliography}{\thebibliography}
\csname @ifundefined\endcsname{endmcitethebibliography}
{\let\endmcitethebibliography\endthebibliography}{}

\end{document}